\documentstyle[epsf]{article}
\begin{document}

\vspace{1.0 cm}

\hfill{ KEK-Preprint~94-99 }

\vspace{-1.0 cm}
\epsfysize3cm
\epsfbox{kekm.epsf}
\begin{center}

{\bf{\Large  Study of a Threshold Cherenkov Counter
             Based on Silica Aerogels with
             Low Refractive Indices
\footnote{to be published in Nucl. Instrum. Meth. {\bf A}} } }

\vspace {1.0cm}

I.~Adachi\footnote{
internet address: ichirou@kekvax.kek.jp
}
, T.~Sumiyoshi, K.~Hayashi,
N.~Iida,
R.~Enomoto, K.~Tsukada \\

{\it
     National Laboratory for High Energy Physics(KEK),
     1-1 Oho, Tsukuba, Ibaraki 305, Japan }

\vspace {0.5cm}

R.~Suda, S.~Matsumoto, K.~Natori

{\it Department of Physics,
     Chuo University,
     1-13-27 Kasuga, Bunkyo-ku, Tokyo 112, Japan }

\vspace {0.5cm}

M.~Yokoyama, H.~Yokogawa

{\it Central Research Laboratory,
     Matsushita Electric Works, Ltd.,
     1048 Kadoma, Osaka 571, Japan }

\end{center}

\begin{abstract}

To identify $\pi^{\pm}$ and $K^{\pm}$ in the region of $1.0\sim 2.5$ GeV/c,
a threshold Cherenkov counter equipped with silica aerogels
has been investigated.
Silica aerogels with a low refractive index of 1.013 have been
successfully produced using a new technique.
By making use of these aerogels as radiators,
we have constructed a Cherenkov counter and
have checked its properties in a test beam.
The obtained results have demonstrated that our aerogel
was transparent enough to
make up for loss of the Cherenkov photon yield due to a low refractive index.
Various configurations for the photon collection system and
some types of photomultipliers, such as the fine-mesh type,
for a read out were also tested.
{}From these studies, our design of a Cherenkov counter dedicated
to $\pi / K$ separation up to a few GeV/c
%in the momentum range of $1.0 \sim 2.5$ GeV/c
with an efficiency greater than $90$ \% was considered.
\end{abstract}
%%%%%%%%%%%%%%%%%%%%%%%%%%%%%%%%%%%%%%%%%%%%%%%%%%%%%%%%%%%%%%%%%%%%%%%%%%%%%%%
%
%
%%%%%%%%%%%%%%%%%%%%%%%%%%%%%%%%%%%%%%%%%%%%%%%%%%%%%%%%%%%%%%%%%%%%%%%%%%%%%%%
%
%==============
\section{ Introduction }
%==============
%
%%%%%%%%%%%%%%%%%%%%%%%%%%%%%%%%%%%%%%%%%%%%%%%%%%%%%%%%%%%%%%%%%%%%%%%%%%%%%%%
%

Recently, asymmetric $e^+e^-$ colliders with high luminosities
to explore $CP$ violation in $B$ meson system( $B$-factory )
have been proposed\cite{kek,slac}.
In a $B$-factory detector, an identification of the particle species
is an important issue. In particular, the separation of $\pi^{\pm}$ and
$K^{\pm}$ in
the momentum region of $1.0 \sim 2.5$ GeV/c is indispensable for
studying many $B$ meson decay channels\cite{kek}.
%, for example, $B\rightarrow DK$\cite{kek}.
%the momentum region of $1.0 \sim 2.5$ GeV/c is indispensable for detecting one
%%of
%the $CP$ violation angles\cite{kek}.
For this purpose, Cherenkov counter equipped with
silica aerogels having a low refractive index is one of the most promising
devices
since it has been widely used
in high energy experiments\cite{carlson}.
Furthermore, the use of an aerogel Cherenkov counter enables us to
reduce the material for a particle identification device
which is located in front of an electromagnetic calorimter.

The requirements imposed on our aerogel Cherenkov counter are:
(i) The refractive index of aerogel( $n$ ) should be in the
$1.010 \sim 1.015$ range in order to to achieve a $\pi$/$K$ separation
capability in the region of $1.0 \sim 2.5$ GeV/c.
(ii) The optical transparency of the aerogel should be
high. Notice that Cherenkov photons
emitted by an injected particle becomes significantly small in our
case because the Cherenkov
light yields are proportional to $1-1/n^2$. Therefore, the loss of photons in
the aerogel
due to absorptions and scatterings should be minimized. (iii) Efficient photon
collection and detection under a strong magnetic field in the counter
have to be considered to avoid any inefficiency for
$\pi$ identification. We require that the detection efficency of $\pi$
identification should be greater
than 90 \%, thus resulting in the averaged number of photoelectrons
greater than 2.5 for $\beta = 1$.

Because aerogels with $n < 1.02$ are not commercially
available, we have developed a new aerogel production technique
with low densities and high transparencies, and a test counter based on
these aerogels has been built.
Using this counter, the optical characteristics of aerogels
and various conditions of counters were tested in order to optimize some
parameters
for our detector design.

In the next section, the production technique of our new aerogel is described.
Their optical properties are given in section 3. Section 4 is devoted
to results using a test beam.
After describing basic properties of the new aerogels,
an investigation of the light collection system,
including tests of photomultipliers, is also given here.
Based on the results from the beam test, a detector consideration is
discussed in section 5.
Finally, section 6 summarizes this paper.

%%%%%%%%%%%%%%%%%%%%%%%%%%%%%%%%%%%%%%%%%%%%%%%%%%%%%%%%%%%%%%%%%%%%%%%%%%%%%%%
%
%==============
\section{ Preparation of Aerogels }
%==============
%
%%%%%%%%%%%%%%%%%%%%%%%%%%%%%%%%%%%%%%%%%%%%%%%%%%%%%%%%%%%%%%%%%%%%%%%%%%%%%%%

\subsection{ Single- and Two-Step Methods }

Silica aerogel with a higher refractive index than $1.02$
has been produced in a conventional
single-step method\cite{tasso:single,kek:single,kawai:single,ins:single}.
In this method,
tetraalkoxysilane and water are mixed in
an excess of alcohol as a solvent.
The following hydrolysis and condensation of silicon alkoxides
with the help of a catalyst are simultaneously proceeded
in a single container:
\begin{eqnarray}
& &
m \mbox{ Si(OR)$_4$ + 4 } m \mbox{ H$_2$O } \nonumber \\
& & \: \: \: \: \: \: \rightarrow
m \mbox{ Si(OH)$_4$ + 4 } m \mbox{ ROH } \\
& &
m \mbox{ Si(OH)$_4$ } \rightarrow
\mbox{ (SiO$_2$)$_m$ + 2 } m \mbox{ H$_2$O }
\end{eqnarray}
In these two successive reactions,
the solution becomes colloidal. Then, aging is done in order to form
three-dimensional networks of SiO$_2$ clusters connected
by a siloxan linkage.
After the alcogel is made, it is dried in an autoclave
at a point safely higher than the critical point( supercritical
drying ) to remove any alcohols.
The density and, consequently the refractive index of the aerogel
is mainly determined from the quantity of alcohol added
in the first stage as a solvent.
If the quantity of alcohol is increased in order to make low density
aerogel, it becomes difficult to form a gel with high porosity
since the backward reactions take place.
This is why aerogels with low refractive
indices or optically good properties can not be made using this method.

To make aerogel with a lower refractive index,
a ``two-step sol-gel'' method has been developed at Lawrence
Livermore National Laboratory\cite{two}. They produce
partially hydrolysed and partially condensed silica oil
in the first step.
Next, the alcohol is distilled off from the silica oil. This
solution is called a ``precursor''.  Then, it is mixed
with a non-alcohol solvent, and polymerization is then performed
by adding an alkali catalyst, which makes SiO$_2$ cluster
pores having a smaller size than an acid catalyst does. The gels, after
aging,
were dried under the supercritical condition, yielding a high-porosity aerogel
with a low refractive index. This two-step method is
one of the best ways to produce aerogels having a low refractive index.
There are, however,
some complicated processes, such as the distilation of alcohol.

\subsection{ New Production Method }

Our new production method of aerogels with a low refractive index is
essentially the same as the single-step method except for the
use of methylalkoxide ``oligomer'' instead of tetraalkoxysilane.
Figs. \ref{fig1} -(a) and -(b) are charts for the
oligomer and the presursor in the two-step method obtained by a gas
chromatograph mass-spectrometer, respectively.
\begin{figure}
\epsfysize8.6cm
\epsfbox{fig1.epsf}
\caption{
Charts for the oligomer(a) and the precursor(b)
measured by the gas chromatograph mass-spectrometer.
The horizontal and vertical directions correspond to
the response time and the content of each composition, respectively.
}
\label{fig1}
\end{figure}
{}From these charts, we can compare compositions of the oligomer with
those of the precursor.
As can be seen,
the compositions of the oligomer are almost identical
to those of the precursor. Namely,
the oligomer is a stable liquid after partial
hydrolysation and partial condensation.
Therefore, the advantage of our method is to simplify
the production processes, which is very important
for the mass-production of aerogels, with keeping
the merits of the two-step method.
In addition,
the oligomer can be easily obtained on a commercial
basis and is not expensive.

Another advantage of our technique is the hydrophobic characteristic
of the aerogels. In the production process, the characteristic of the
alcogel can be changed so as to be hydrophobic by replacing any
silanol groups( -OH ) into hydrophobic ones.

Our production methods are as follows.
At first, two solutions are prepared. One is made by adding an oligomer
in ethyl alcohol as a solvent, and the other is distilled water with
ethyl alcohol.
Here, ammonia water is also added in this solution
as an alkali catalyst to polymerize small and uniform pores.
These two solutions are mixed in a polystyrene container at room temperature.
After a few minutes, the gelation takes place, and the alcogel is made.
The alcogel is put into a sealed vessel in order to
suppress evaporation of the ethyl alcohol, and is kept at
room temperature for one week.
Then, the sealed vessel, in which the alcogel has been placed,
is filled with ethyl alcohol vapor in order to proceed
the formation of a three-dimensional network of SiO$_2$.
After that, alcogel is moved into a stainless-steel bag from
a polystyrene container and is left in ethyl alcohol for one week.

To make
alcogel hydrophobic, special treatment is taken.
The silanol groups on the surface of SiO$_2$
is likely to be charged and contacting other ions.
It is thus replaced into Si(CH$_3$)$_3$ groups by
adding hydrophobic agent, for example, hexamethyldisilazane,
into ethyl alcohol, as shown in Fig. \ref{fig2}.
\begin{figure}
\epsfysize12.6cm
\epsfbox{fig2.epsf}
\caption{
Schematic drawing of the hydrophobic reaction of the alcogel.
}
\label{fig2}
\end{figure}
This solution, in which the alcogel is moved,
is stirred by an electrical power agitator at
$40\sim60^\circ$. This process
is completed in $3\sim24$ hours, which depends on the temperature.
The replacement of the hydrophilic groups in the alcogel
with hydrophobic ones also plays an important role in preserving
its volume after supercritical drying in an autoclave since
shrinkage of the aerogel is considered to be due to
the electrically attractive force among the silanol groups
and/or between the silanol groups and other ions.

After this reaction, the alcogel is again sunk in ethyl alcohol, and
the alcohol is replaced once every 3 days. During this period,
ammonia in the alcogel, which was generated in the hydrophobic
reaction, is solved into alcohol.

To transform the alcogel into aerogel,
the alcogel is dried at the supercritical point.
Here, alcohol is replaced into carbon dioxide before heating.
The pressure and temperature at the supercritical point of carbon dioxide is
much
lower than that of ethyl alcohol. It, therefore, provides a safer and
more economical benefits.
A complete description can be found in Ref.\cite{matsushita}

\begin{figure}
\epsfysize8.1cm
\epsfbox{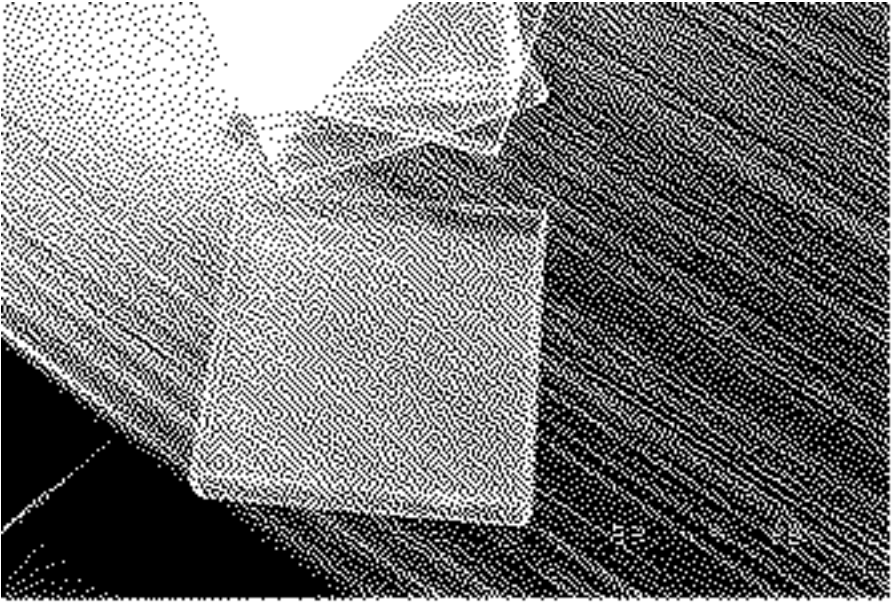}
\caption{
Photograph of the aerogel produced using the new technique.
}
\label{fig3}
\end{figure}

An aerogel of $16\times 16 \mbox{(cross section)} \times 2 \mbox{(thick)}$
cm$^3$
with $n = 1.013$ has been successfully made using this method.
We produced 22 pieces of aerogels with this size.
Fig. \ref{fig3} is a photograph of the obtained aerogel.
It was shrunk to be about $20$ \% in volume.
Some cracks could be
seen inside the volume or on its surface although
they were tolerable.
Its optical properties were described in the next section.
%Optimizations of the conditions, especially in the
%supercritical drying, are now in progress to prevent from
%shrinkages.

%%%%%%%%%%%%%%%%%%%%%%%%%%%%%%%%%%%%%%%%%%%%%%%%%%%%%%%%%%%%%%%%%%%%%%%%%%%%%%%
%
%==============
\section{ Optical Properties of New Aerogels }
%==============
%
%%%%%%%%%%%%%%%%%%%%%%%%%%%%%%%%%%%%%%%%%%%%%%%%%%%%%%%%%%%%%%%%%%%%%%%%%%%%%%%

\subsection{ Transmission }

\begin{figure}
\epsfysize18cm
\epsfbox{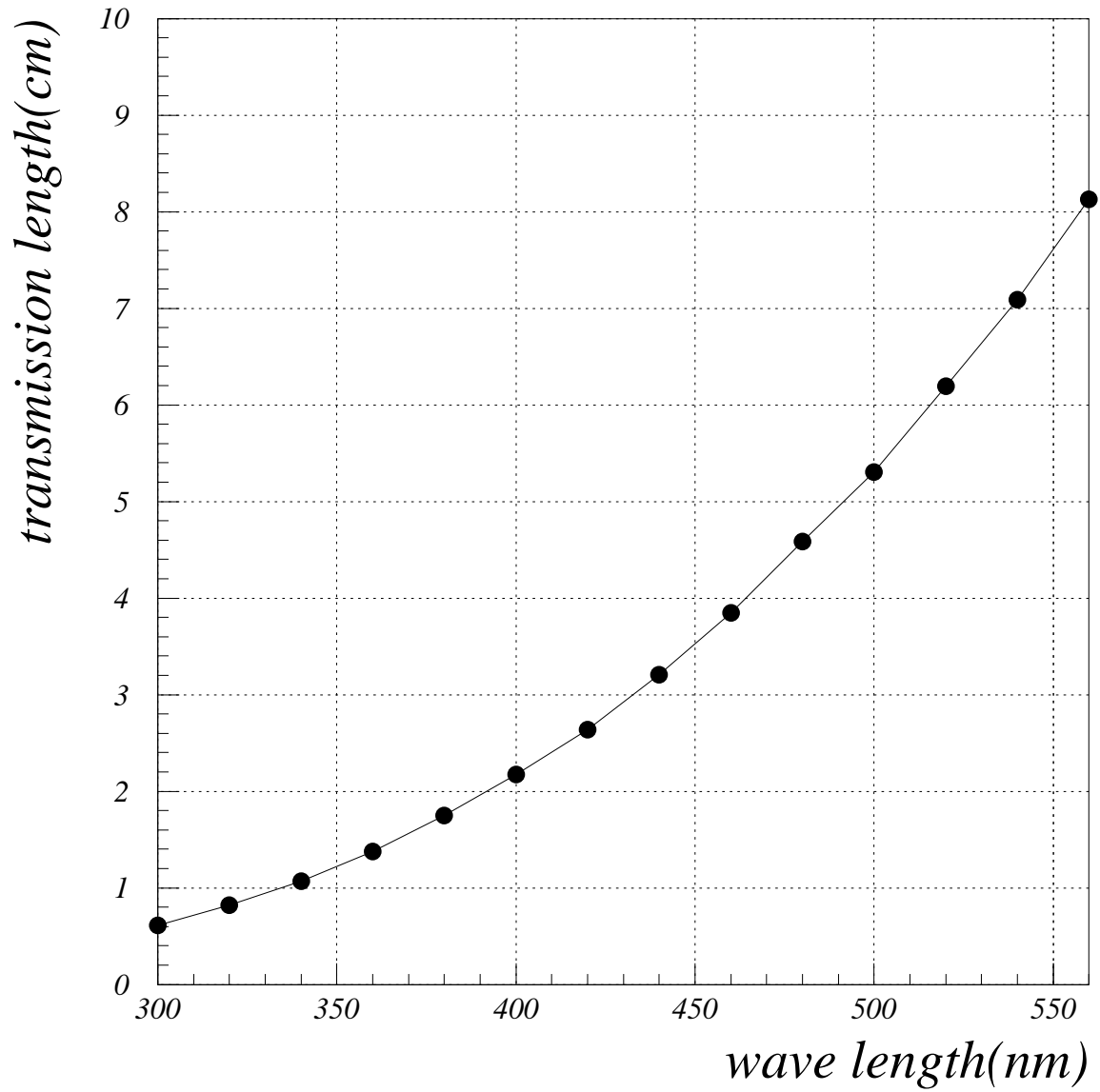}
\caption{
Transmission length( cm ) as a function of the wave length( nm )
obtained with a photospectrometer. The aerogel thickness is 2.2 cm.
}
\label{fig4}
\end{figure}

The transmission of the aerogel obtained with the new method described
above was measured using a photospectrometer\cite{spectr}.
The thickness of
the aerogel of the sample was 2.2 cm. The obtained data can be translated into
the transmission length( $\Lambda$ ) by the function:
\begin{equation}
T = T_0 \exp( - d / \Lambda ),
\end{equation}
where $T$ and $d$ are the transmission and thickness of the
aerogel, respectively.
Fig. \ref{fig4} shows the
transmission length as a function of the wave length( $\lambda$ ).
The transmission length strongly depends on the
wave length.
%%%%%
%In the ultraviolet region, $T$ was governed by
%absorption effects while the Rayleigh scattering dominated in the
%visible-light region.
%%%%%
These effects can be related to
the size and uniformity of microscopic pores of the aerogel.
Our results are compared with those of other measurements in
Table~\ref{table:trans}.
\begin{table}
\begin{center}
\begin{tabular}{cccccc}    \hline\hline
wave length       & this experiment &     Ref.\cite{tasso:single} &
            Ref.\cite{kek:single}   &     Ref.\cite{kawai:single} &
                                          Ref.\cite{ins:single}
\\
 ( nm ) & $n=1.013$ & $n=1.024$ & $n=1.058$ & $n=1.058$ & $n=1.055$  \\ \hline
 400    &       2.2 &           &       1.5 &     1.38  &     1.15   \\
 440    &       3.2 &      2.64 &       2.0 &           &            \\
\hline\hline
\end{tabular}
\end{center}
\caption{ Summary of measurements of the transmission length( cm ). }
\label{table:trans}
\end{table}
Typically, the transmission length was obtained to be more than 3.0 cm at
$\lambda =
440$ nm.
Compared with aerogels used in high energy experiments so far,
our aerogel is one of the most transparent ones.

\subsection{ Refractive Index }

The refractive indices were determined using an optical method with a
He-Ne laser beam of $\lambda = 632.8$ nm before the beam test.
The minimum deflection of
the incident laser beam after transversing one corner of the aerogel,
which was placed on a movable table, was measured.
This measurement was made for all of the aerogels produced.
Most of them had refractive indices of $1.013$,
suggesting uniformity of the optical quality of the material.

%%%%%%%%%%%%%%%%%%%%%%%%%%%%%%%%%%%%%%%%%%%%%%%%%%%%%%%%%%%%%%%%%%%%%%%%%%%%%%%
%
%==============
\section{ Results of Beam Test }
%==============
%
%%%%%%%%%%%%%%%%%%%%%%%%%%%%%%%%%%%%%%%%%%%%%%%%%%%%%%%%%%%%%%%%%%%%%%%%%%%%%%%

\subsection{ Experimental Set-up }

This test was carried out at the $\pi2$ beam line in KEK $12$-GeV/c
PS. The experimental layout is shown in Fig. \ref{fig5}.
\begin{figure}
\epsfysize8.1cm
\epsfbox{fig5.epsf}
\caption{
Schematic drawing of the experimental set-up.
}
\label{fig5}
\end{figure}
Negatively charged
particles, mainly $\pi^-$, in the momentum range of $0.6 \sim 4.0$ GeV/c
were used.
The contribution from $K^-$ was sufficiently small so as to be neglected
over this momentum region. Although the fraction of
electrons was small at $4.0$ GeV/c, it increased with lower momentum
and was sizeable below $1.0$ GeV/c.

A test counter was located inside a tight black box for light shield.
Two gas Cherenkov counters( $C1$ and $C2$ ) were used to remove
electrons from the pion beam.
Three scintillation counters( $S1 \sim S3$ ) and one anti-counter( $AC$ )
were placed upstream of the test counter, and one scintillation
counter( $S4$ ) was located downstream.
$S1$ and $S2$ have cross sections of $5 \times 5$ cm$^2$ and $2 \times 2$
cm$^2$,
respectively. The beam was defined to be $1\times1$ cm$^2$ by the $S3$ counter.
$AC$ was placed in front of the test
counter to make sure that only one charged particle hits the test counter.
The beam direction was definitely guaranteed by the $S4$ counter.
A trigger was generated by the 3-fold coincidence of $S1 \sim
S3$.

Analog signals from the test counter, together with all other counters,
were fed into an ADC Lecroy 2249W, which was driven via a CAMAC system
controlled from a personal computer.

In the off-line analysis, events which left reasonable signals
in all of $S1 \sim S4$ counters and simultaneously no signals in $AC$ were
used.
The electron contamination was also rejected by the $C1$ and $C2$ informations.

\subsection{  Test Counter Configuration }

Fig. \ref{fig6} shows a schematic drawing of the test Cherenkov counter.
\begin{figure}
\epsfysize14cm
\epsfbox{fig6.epsf}
\caption{
Schematic view of the test Cherenkov counter.
The $x$- and $y$-axes were also indicated.
The origin( $(x,y) = (0,0)$ ) was defined to be the center of the counter.
}
\label{fig6}
\end{figure}
The test Cherenkov counter has a size of $16 \times 16$ cm$^2$ cross
section and $25$ cm height, where the active aerogel volume is $16 \times
16 \times 14$ cm$^3$. Two holes with $2''$ diameters were made in
the side walls of the box to place photomultipliers( PMT's ).
Aerogel sheets were stacked in the box, and $2''$ PMT's
were equipped in the holes for read out as indicated in Fig. \ref{fig6}.
The internal surface was covered with white reflectors of Millipore
sheets( GVHP00010 ).

To check the optical characteristics of the aerogels, $2''$ quantacon-type PMT
( R3241 provided by Hamamatsu P.K.K.~),
which is able to detect a single photoelectron, was used.
The high voltages applied to two PMT's were determined so that two gains were
identical. This adjustment was carried out with a stable LED light source,
of which the intensity was reduced to be equivalent to a single photon.

\subsection{ Evaluation of the Average Number of Photoelectrons }

\begin{figure}
\epsfysize9cm
\epsfbox{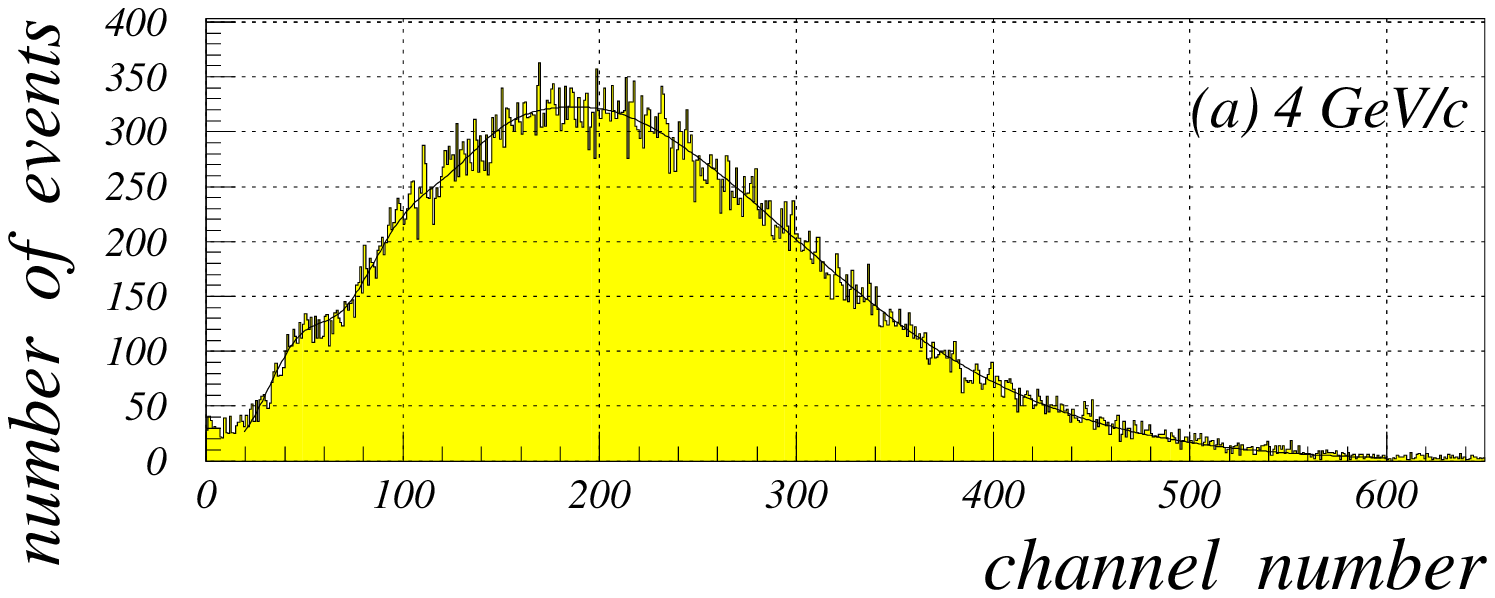}
\epsfysize9cm
\epsfbox{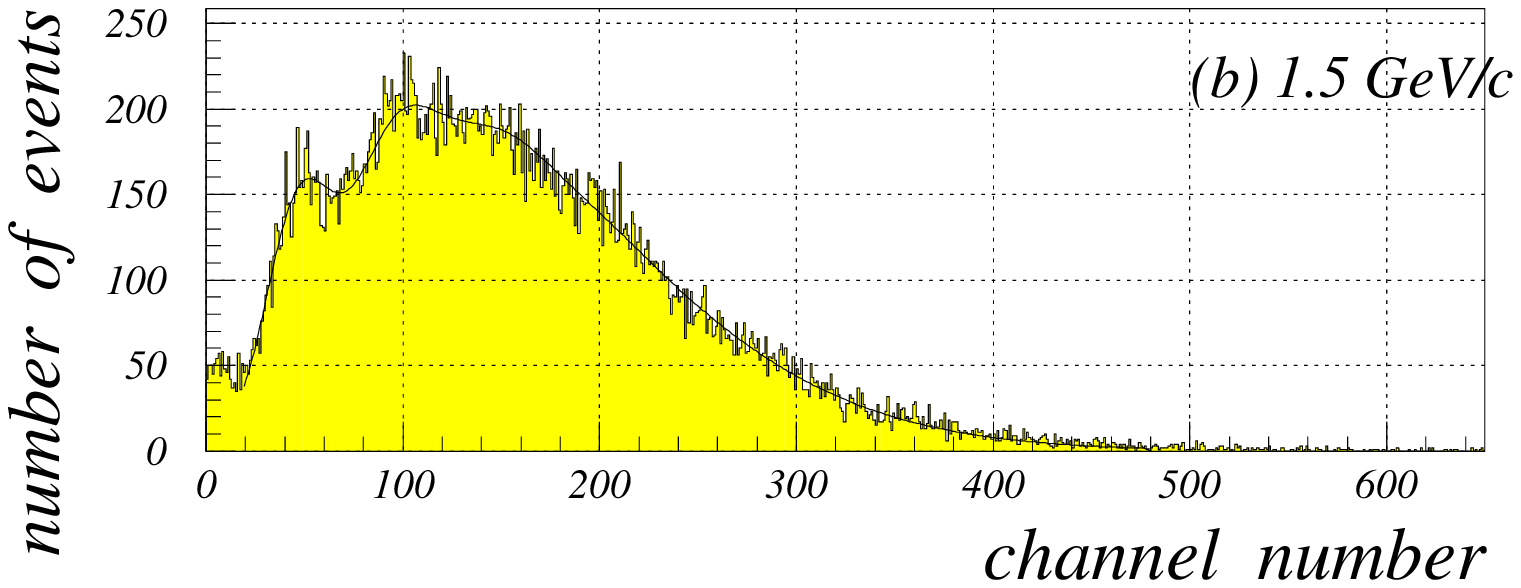}
\caption{
Pulse height distribution obtained by summing two $2''$
quantacon-type PMT's
for (a) 4 GeV/c and (b) 1.5 GeV/c $\pi$ beam.
The hatched histograms indicate the obtained data, whereas the solid
curve is the best fit.
}
\label{fig7}
\end{figure}

Figs. \ref{fig7} -(a) and -(b) show the pulse height spectrum obtained
by summing two $2''$ PMT's
for $4$ GeV/c and $1.5$ GeV/c $\pi^-$, respectively.
We extracted an average number of photoelectrons( $N_{pe}$ ) from these
distributions.
Some methods have been adopted to determine $N_{pe}$.

One method is to calculate the inefficiency from the number of
pedestal events( $N_{ped}$ ). Assuming a Poissonian distribution, one can
determine $N_{pe}$ as:
\begin{equation}
N_{pe} = - \log( N_{ped} / N_{all} ),
\label{eq:pedestal}
\end{equation}
where $N_{all}$ is the total number of events. In this method,
$N_{pe}$ can be evaluated without ambiguity when $N_{pe}$ is less
than 5.
If $N_{pe}$ becomes more than 5, we can not easily separate $N_{ped}$
from the over-all pulse height distribution.

Another method is to fit the distribution by the function of the convolution of
the Gaussian and Poissonian distributions expressed by:
\begin{equation}
f(x) = C  \sum^{k_0}_{k=1} \frac{ ( N_{pe} )^k  e^{ -N_{pe} } }{k !}
 	\cdot \frac{1} { \sigma \sqrt{ 2\pi k } }
        \exp ( \frac{ - ( x - p k )^2} { 2k\sigma^2 } ),
\label{eq:fit}
\end{equation}
where $p$ corresponds to the interval between zero and single
photoelectron peaks. $\sigma$ is the standard deviation of a single
photoelectron peak, and $x$ is ADC channel after pedestal subtraction.
$C$ is a normalization constant. $k_0$ is the maximum number of
photoelectron peaks contained in the distribution.
This function assumed that
the $k$-th photoelectron peak has a standard deviation of
$\sqrt{k}\sigma$ and that the interval of photoelectrons is the same.
In the fitting procedure, 4 parameters( $N_{pe}$, $\sigma$, $p$ and
$C$ ) were allowed to be free, whereas $k_0$ was fixed to be 10 so that
the function of eq.\ref{eq:fit}~could well reproduce the whole shape
of the distribution.

In this analysis,
the second method was employed unless specified.
As a cross check, the first method was also used for several measurements, and
the results from the two methods agreed well.

\subsection{ Thickness }

\begin{figure}
\epsfysize9cm
\epsfbox{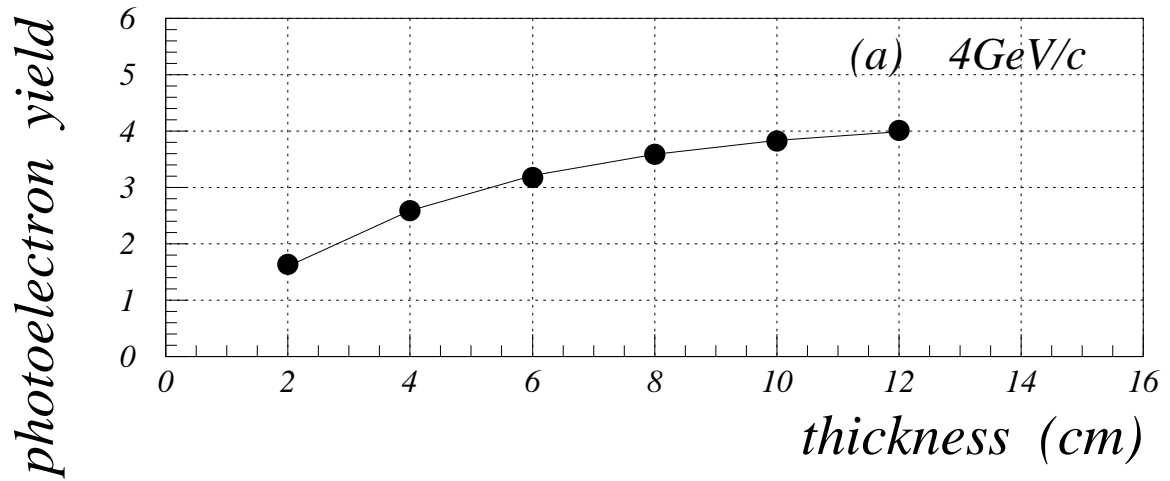}
\epsfysize9cm
\epsfbox{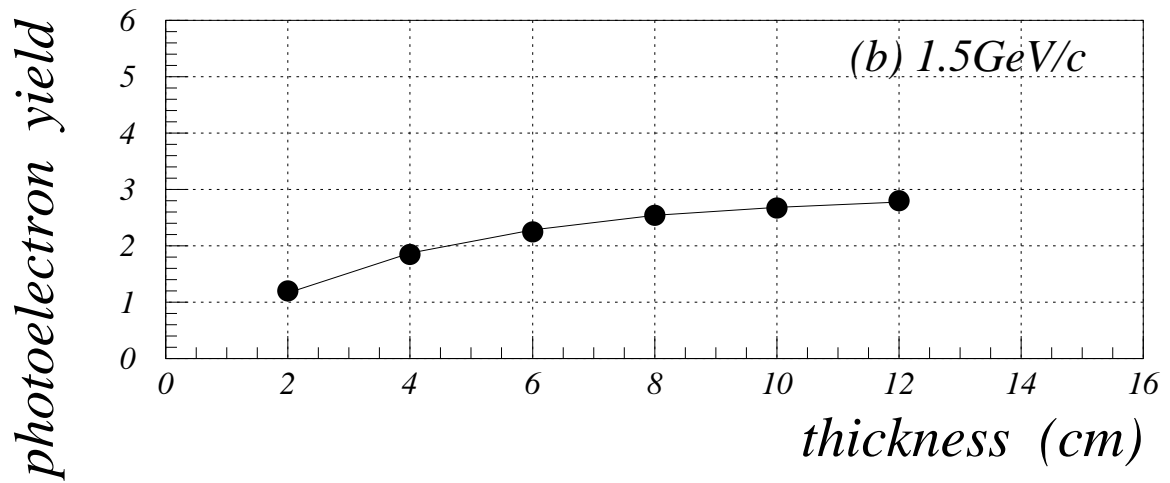}
\caption{
Average number of photoelectrons as functions of the aerogel
thickness for (a) 4 GeV/c and (b) 1.5 GeV/c $\pi$ beam.
The measured data are plotted as closed circles, and the solid curves
show the best-fitted function.
}
\label{fig8}
\end{figure}

The first check was made in order to determine the aerogel thickness to be used
as
a radiator. The thickness was varied by installing an additional wall,
of which both sides were covered with Millipore sheets,
between aerogel sheets so that the distance between the surface of the
aerogel and the PMT was maintained. The beam was injected at the center of
the test counter.
Figs. \ref{fig8} -(a) and -(b) show $N_{pe}$ as functions of the thickness for
$1.5$ GeV/c and
$4$ GeV/c $\pi^-$, respectively.
These behaviors can be well parametrized by:
\begin{equation}
N_{pe} = N_{max} ( 1 - \exp( -d/ L ) ),
\end{equation}
where $d$ is the thickness, and $L$ stands for the effective
absorption length. $L$ was obtained to be $4.23\pm 0.01$ cm and
$3.92\pm 0.04$ cm for
$4.0$ GeV/c and $1.5$ GeV/c, respectively. $N_{max}$ is $4.23\pm 0.03$
for $4.0$ GeV/c $\pi$.
The best-fit curves are also shown in the figures.
The fitted curves well follow the obtained data.
The photoelectron yield saturates at $d \sim 12$ cm, and
$94.2$ \% of the photoelectrons can be detected at $d = 12$ cm.

In addition, we placed one more aerogel sheet into the counter, resulting
in a thickness of $14$ cm. However, the distance between the surface of
the aerogel and the PMT's becomes smaller. The measured $N_{pe}$
was increased to be $4.57\pm0.01$ by this configuration.
This value is quite satisfactory.
Though the Cherenkov photon yield significantly decreases as
$1 - 1 / n^2$ because of the low
refractive index, we can compensate
for this loss by producing highly transparent aerogels.
%High transparency of our aerogel can compensate loss
%of Cherenkov photons from low refractive index.

In what follows, we used this set-up having $14$ cm
aerogel sheets as a reference configuration.

\subsection{ Threshold Behavior }

\begin{figure}
\epsfysize18cm
\epsfbox{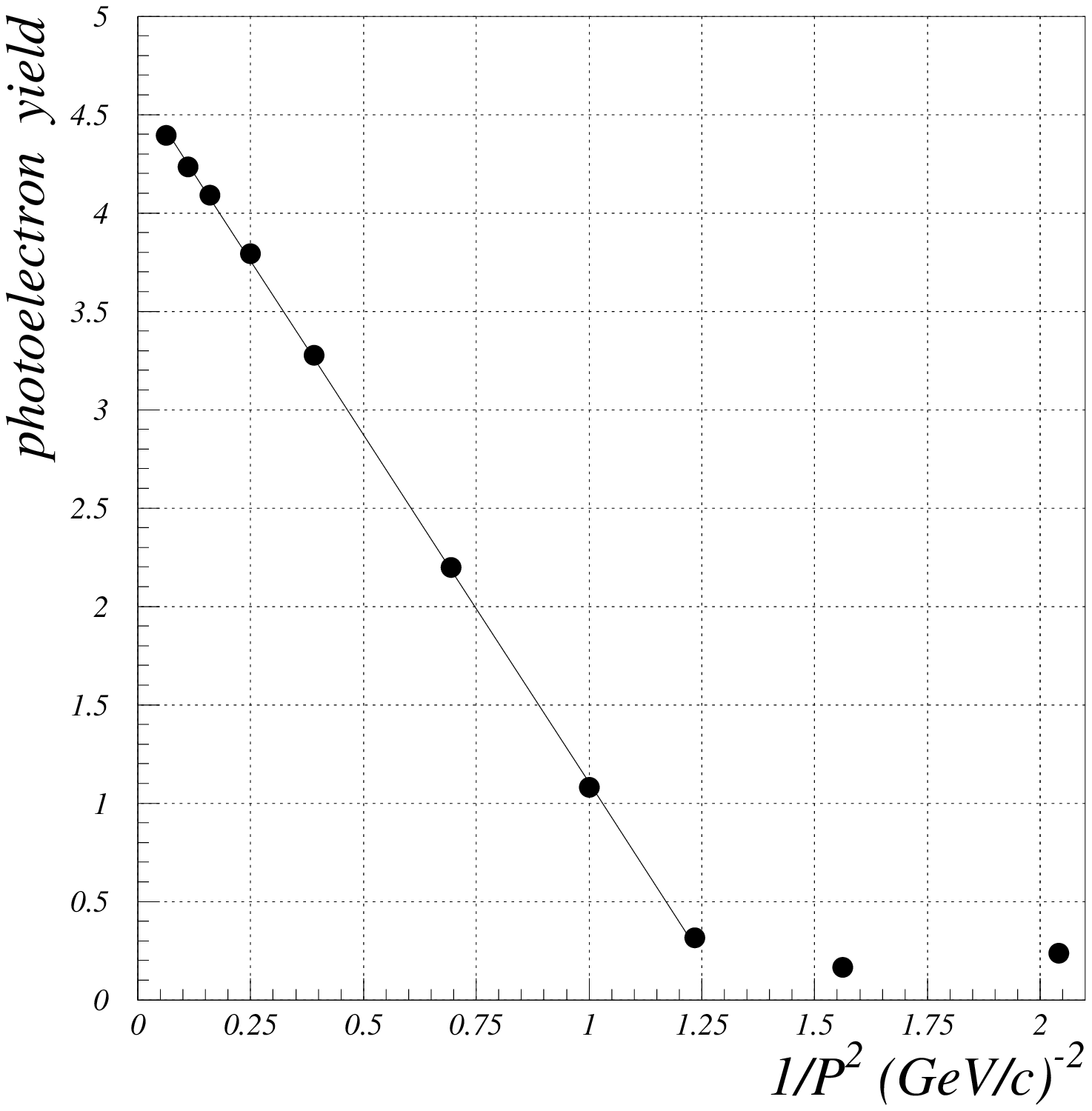}
\caption{
Average photoelectron yields as a function of the inverse momentum squared
( (GeV/c)$^{-2}$ ). The closed circle is the obtained data, and the
straight line is a fit to the data. The refractive index was obtained
to be $n=1.0127 \pm 0.0001$ from this fit.
}
\label{fig9}
\end{figure}

To measure the refractive indices of aerogels, the photoelectron yield
was obtained by changing
the momentum of the incident $\pi^-$ particle which was injected
at the center of the aerogel area. $N_{pe}$ as a function of the inverse-square
momentum of the incident particles is plotted in Fig. \ref{fig9}.
In this analysis, $N_{pe}$ was
deduced from the pedestal events( eq.~\ref{eq:pedestal} )
in the region where the photoelectron yield was less than 1,
since only pedestal peak can be seen in the obtained pulse height distribution
and thus it is difficult to fit the function of the form of eq.~\ref{eq:fit}.
The momentum dependence of $N_{pe}$ can be
given by:
\begin{equation}
N_{pe} = C ( 1 - \frac{1}{n^2} ( 1 + \frac{m^2}{p^2} ) ),
\end{equation}
where $n$ is the refractive index of aerogel, and $m$ and $p$ denote the mass
and
momentum of the incident particle, respectively.
$C$ is a constant which corresponds to the
number of photoelectrons obtained at $\beta = 1$.
The refractive index was estimated to be $1.0127\pm 0.0001$
by fitting the obtained data above
the threshold. Therefore, the threshold momentum was obtained to be
$0.863$ GeV/c and $3.05$ GeV/c for $\pi^{\pm}$ and $K^{\pm}$, respectively.
This value was quite consistent with that from the optical measurement
described previously.

It is important to estimate the knock-on effect because this effect
produces a signal even for an incident particle with momentum below the
threshold, causing a miss-identification of $K$ as $\pi$.
The knock-on effect can be estimated to be $10$ \%
from the photoelectron value below the threshold momentum. This value
is comparable with that obtained by other measurements\cite{tasso:single}.

\subsection{ Tests of The Counter Configuration }

Some configurations for Cherenkov photon collection were tested.
Basically, there are two kinds of photon collection systems.

One is
a diffuse scattering technique, in which Cherenkov photons undergo
scattering at the walls, which are covered with a highly efficient reflector,
and
reach the window of the PMT's. Accordingly, the directed photons as well as
the scattered photons have a chance of being detected. We used this system as a
reference configuration( as described before ). Here, our interest is in the
position dependence of the PMT's. Thus, the positions of the read out
PMT's were changed as sketched in Fig. \ref{fig10}-(a) ( type A ).
\begin{figure}
\epsfysize17.3cm
\epsfbox{fig10.epsf}
\caption{
Schematic drawing of the test counter for tests of the light
collection systems. The $x$- and $y$-directions are defined in the figure.
The center of the counter was taken as the origin( $(x,y) = (0,0)$ ).
(a) The diffuse scattering system.
Two PMT's are faced each other( type A ).
(b) The mirror scattering system( type B ). The mirror is made of thin
aluminum, of which the surface is covered with MgF$_2$.
}
\label{fig10}
\end{figure}

Another type is to use a mirror to collect Cherenkov photons.
This method collects only directed photons, which are determined by
mirror optics. This type was also tested by installing a flat mirror as shown
in Fig. \ref{fig10}-(b) ( type B ). The mirror was made of thin aluminum
evaporated onto the fused glass. The surface of the aluminum was coated with
MgF$_2$.

In the case of center injection at $4.0$ and $1.5$ GeV/c $\pi^-$,
the results for $N_{pe}$ are listed in Table~\ref{table:config}.
\begin{table}
\begin{center}
\begin{tabular}{ccccc}   \hline\hline
momentum           & \multicolumn{2}{c}{reference config.} &  type A
   & type B
\\ \cline{2-5}
\hfill ( GeV/c )
                   & quantacon-type & fine-mesh     & quantacon-type &
quantacon-type
\\ \hline
 1.5               &  $3.18\pm0.01$ & $1.95\pm0.05$ & $3.53\pm0.01 $ &
$3.96\pm0.13$
\\
 4.0               &  $4.57\pm0.01$ & $2.74\pm0.08$ & $4.70\pm0.01 $ &
$4.77\pm0.17$
\\ \hline\hline
\end{tabular}
\end{center}
\caption{ Average number of photoelectrons obtained from various
configurations. The types of the $2''$photomultipliers used are also
listed.
$\pi^-$ was injected at the center of the aerogel. }
\label{table:config}
\end{table}

The position dependence of the incident particle was measured for
three counter configurations; basic configuration( for a reference ),
type A and type B. The beam used was $4.0$ GeV/c $\pi^-$.
The $x$ and $y$ directions were defined as indicated in Fig. \ref{fig10}.
Fig. ref{fig11} shows $N_{pe}$ as a function of $y$ at different $x$ for the
reference configuration.
\begin{figure}
\epsfysize18cm
\epsfbox{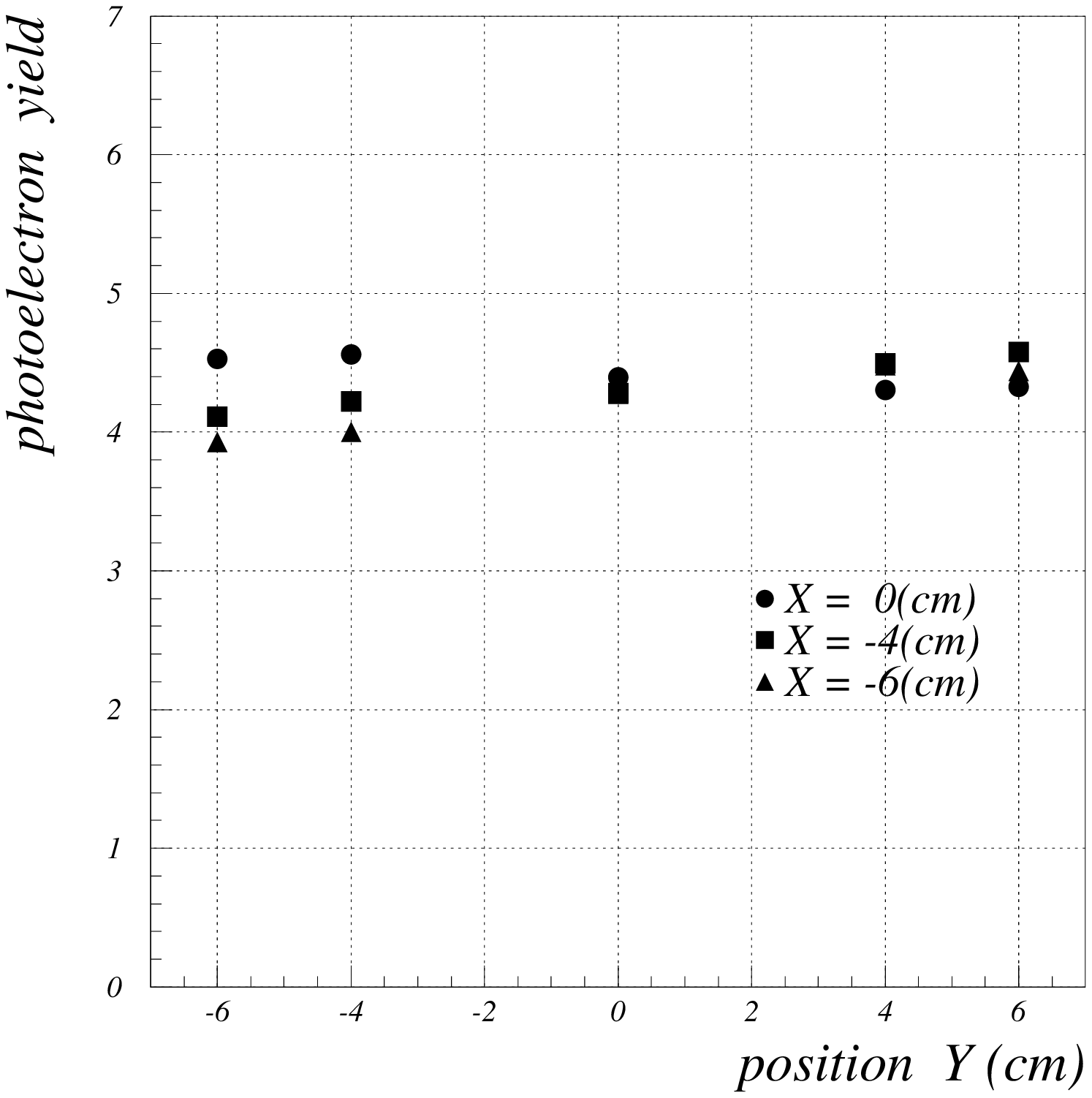}
\caption{
Average number of photoelectrons as a function of $y$ ( cm ) at
different $x$ ( cm )for a reference configuration. 4.0 GeV/c $\pi$ was
used.
}
\label{fig11}
\end{figure}
After summing pulse heights of two PMT's,
the resultant $N_{pe}$ has almost uniform response. The same figure, but
for type A, is shown in Fig. \ref{fig12}.
\begin{figure}
\epsfysize18cm
\epsfbox{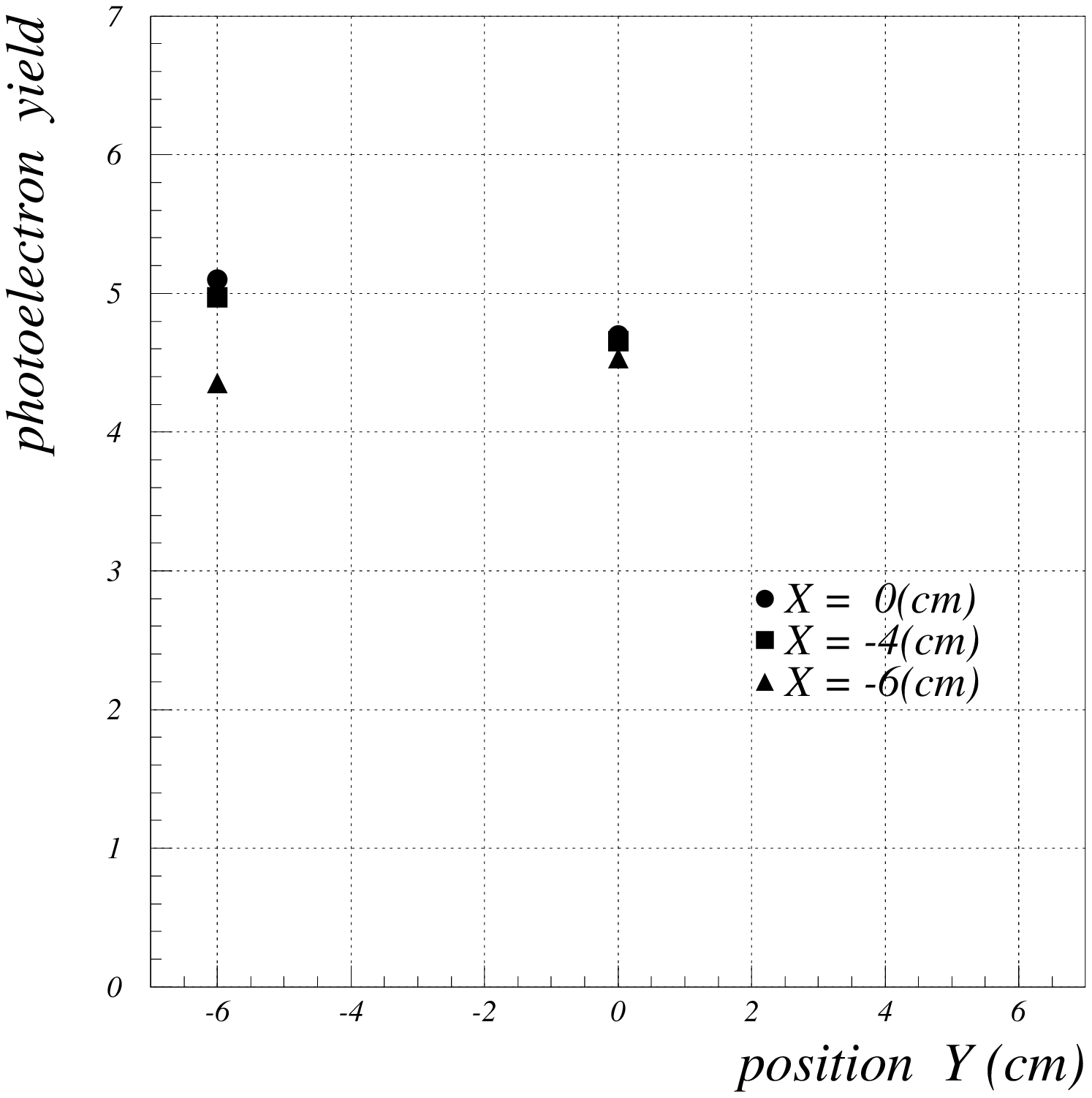}
\caption{
Same distribution of Fig.11, but for type A.
}
\label{fig12}
\end{figure}
The same tendency can be seen.
The results for type B are shown in Fig. \ref{fig13}.
\begin{figure}
\epsfysize18cm
\epsfbox{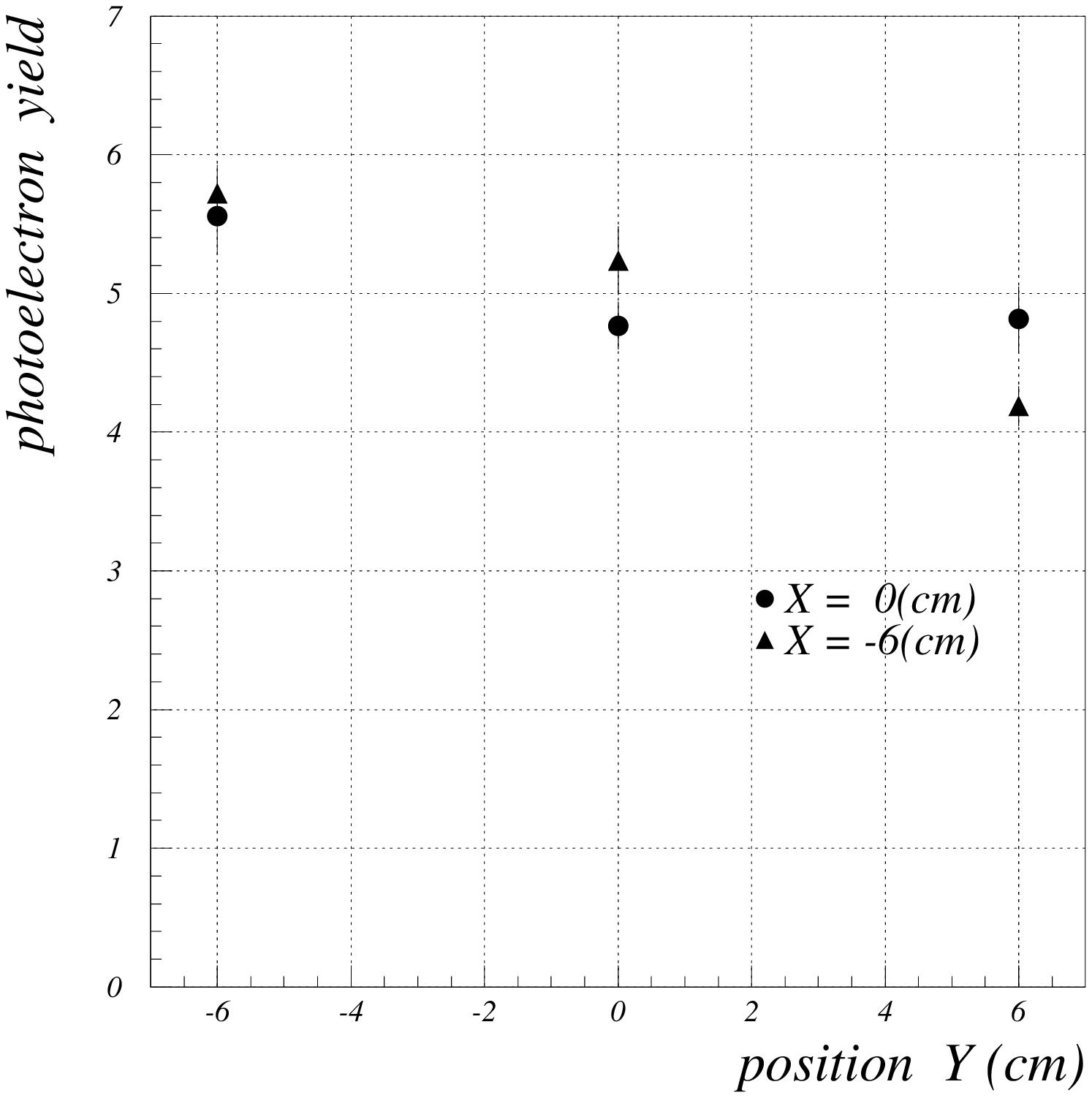}
\caption{
Same distribution of Fig.11, but for type B.
}
\label{fig13}
\end{figure}
Although the mirror option provides the largest $N_{pe}$, it
causes a strong dependence of the beam position.
%, which comes from the
%optics used, which selects the directionality of Cherenkov photons.

{}From this study,
it was found that type A gives better results than those from the
reference configuration at the center injection and that for the uniform
position dependence,
on the other hand, type B provides maximum $N_{pe}$, but
has a strong dependence on the beam position.
Therefore, type A will be adopted for our final design.

\subsection{ Tests of Photomultipliers }

\begin{figure}
\epsfysize18cm
\epsfbox{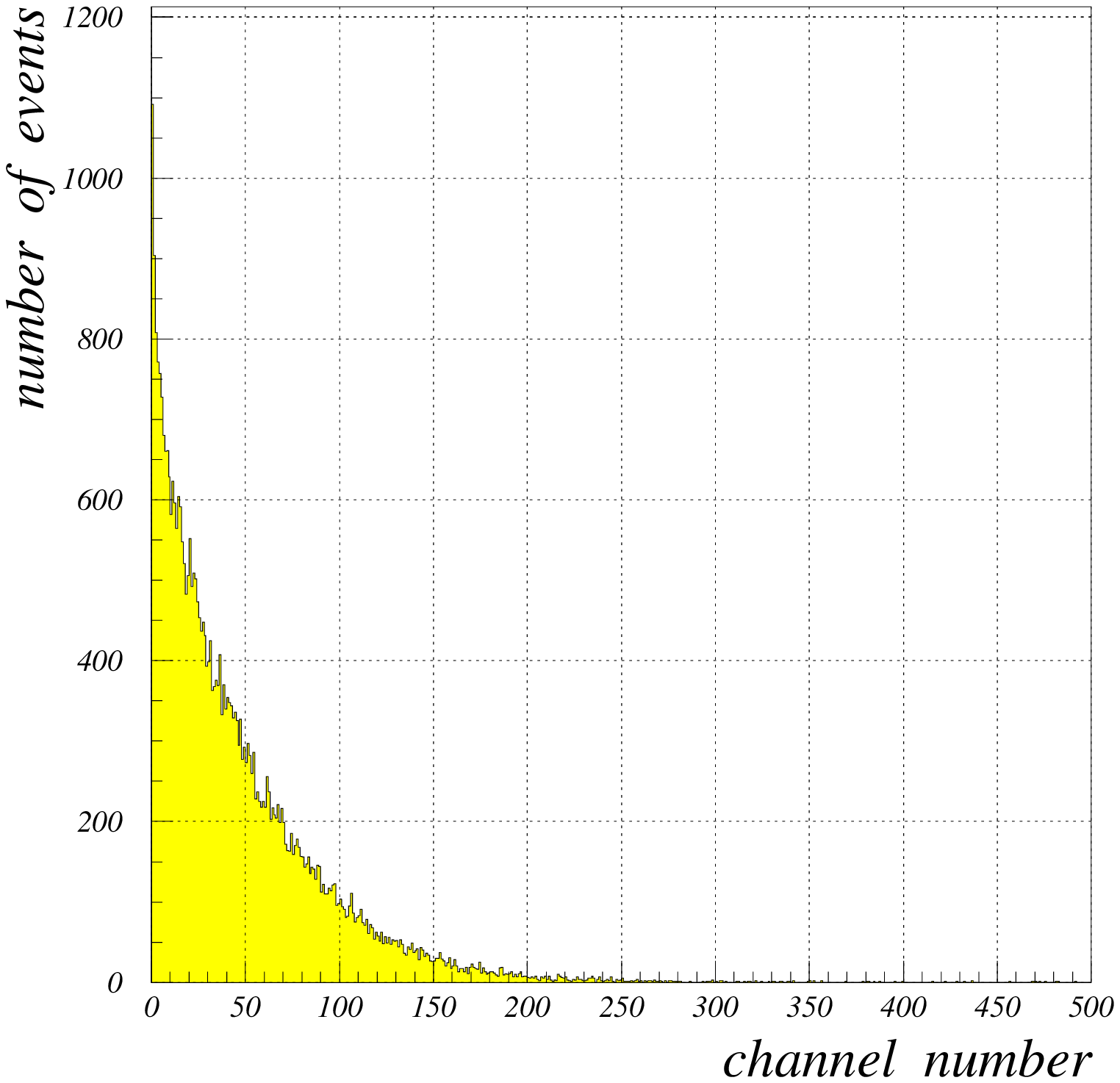}
\caption{
Pulse height distribution from one $2''$ fine-mesh PMT at 4.0 GeV/c $\pi$.
}
\label{fig14}
\end{figure}

In a realistic environment, a read out for an aerogel Cherenkov counter should
be
done in a strong magnetic field of 1 tesla. For this purpose, a fine-mesh type
photomultiplier has been developed. To check the feasibility for aerogel read
out
with fine-mesh PMT's, two $2''$ quantacon-type PMT's were replaced into two
$2''$ fine-mesh PMT's( R2490-05, assembly type, H2611
manufactured by Hamamatsu P.K.K.~).
This PMT contains 16 dynodes in the amplification stage.
The gains of two PMT's were adjusted
in the same way as mentioned previously.
Here, a reference counter configuration was used, and $4.0$ GeV/c and
$1.5$ GeV/c
$\pi$ beams were injected. Fig. \ref{fig14} indicates the pulse height
distribution detected
from one fine-mesh PMT.
No peak, but a long exponential-like tail, was observed.
This phenomena is described in Ref.\cite{finemesh}.
$N_{pe}$ was extracted by counting the
number of pedestal events. However, there was an ambiguity in
separating pedestal events
from the signal distribution. This uncertainty was estimated by
varying the threshold
value by which the pedestal was discriminated, and was obtained to be $10$ \%.
$N_{pe}$ at $4.0$ GeV/c was calculated to be $2.74\pm0.08$ as listed in
Table~\ref{table:config}.
This value is about $60$ \% smaller than that from
quantacon-type PMT's. The difference is considered to result from the sensitive
area
of a photocathode mounted in fine-mesh PMT's.
%This value of $N_{pe}$ can be
%improved by; (1) producing high transparent aerogels, (2) changing refractive
%%index
%from $1.013$ to $1.018$ to increase intrinsic Cherenkov photon yields.
%These studies are now in progress.

%%%%%%%%%%%%%%%%%%%%%%%%%%%%%%%%%%%%%%%%%%%%%%%%%%%%%%%%%%%%%%%%%%%%%%%%%%%%%%%
%
%==============
\section{ Detector Consideration }
%==============
%
%%%%%%%%%%%%%%%%%%%%%%%%%%%%%%%%%%%%%%%%%%%%%%%%%%%%%%%%%%%%%%%%%%%%%%%%%%%%%%%

Based on these tests, we found that our aerogel has good
characteristics. Light collection of type A( described in section 4 ) will be
optimal since it gives reasonable number of photoelectrons and small
position dependence.

The use of fine-mesh PMT's for the counter read out decreases
the photoelectron yields.
Nevertheless,
the number of photoelectrons obtained from two $2''$ fine-mesh PMT's
( $2.74\pm0.08$ ) is sufficiently large for a negative identification of
$K^{\pm}$,
%which is important to separete $B \rightarrow \pi^+\pi^-$ from the
%background reaction $B \rightarrow \pi^{\pm} K^{\mp}$ since these
%branching ratios are considered to be similar.
However, it does not work as a positive identification of $K^{\pm}$ out of a
huge number
of $\pi^{\pm}$. In the case of $B^{\pm} \rightarrow D^0 K^{\pm}$, for
example, this feature is crucial\cite{kek}.

According to these considerations,
an increase in the photoelectron yields is particularly necessary.
It can be improved by; (i) changing the refractive index from 1.013 to
1.018. By making this change, we can gain intrinsic Cherenkov photons by a
factor of 1.4. Simultaneously, however, this trades the particle
separation capability. The threshold momentum for $K$ is reduced to be
2.59 GeV/c from 3.05 GeV/c.
(ii) producing more transparent
aerogels by optimizing some conditions in the production processes such as
temprature.
More transparent aerogels can increase the effective absorption length
of the counter. This allows us to stack more aerogel sheets in the
counter box.

Moreover,
there are still some cracks in our aerogel as described in section 2.
Thus, the optimizations of some parameters such as pressure
in the supercritical drying should be done.
%These works for improvements are required.

%%%%%%%%%%%%%%%%%%%%%%%%%%%%%%%%%%%%%%%%%%%%%%%%%%%%%%%%%%%%%%%%%%%%%%%%%%%%%%%
%
%==============
\section{ Conclusion }
%==============
%
%%%%%%%%%%%%%%%%%%%%%%%%%%%%%%%%%%%%%%%%%%%%%%%%%%%%%%%%%%%%%%%%%%%%%%%%%%%%%%%

Silica aerogels having refractive indices of $1.013$ have been successfully
produced. A Cherenkov counter based on new aerogels was constructed, and
its properties were studied with the test beam. The optical properties of
aerogels
were good enough to compensate for the intrinsically small number of
photoelectrons
emitted by a charged particle.

Two types of light collection systems, diffuse and mirror reflection, were
tested.
The mirror option provided a large number of photoelectrons, however,
strong position dependence of the injection point was observed. The diffuse
scattering
option, on the other hand, gave reasonable results.

Moreover, fine-mesh photomultipliers, which are able to detect photons
under a strong magnetic field, were used for the read out. The resultant light
yields were marginally feasible for our experiment.

Based on these studies, our design of a Cherenkov counter, which can identify
$\pi$ and $K$ in the region of $1.0 \sim 2.5$ GeV/c, was considered.
%We will make use of the diffuse light collection method.
To increase the number of
photoelectrons observed with fine-mesh photomultipliers, the refractive index
of
silica aerogels should be increased to be $n = 1.018$.

%\vspace{2.0cm}

\section*{Acknowledgement}

This work was partially supported by a collaborative research program
between Matsushita Electric Works, Ltd. and KEK.
We are indebted to S.~Iwata, F.~Takasaki and M.~Kobayashi at KEK for their
continuous encouragement.
We are grateful to H.~Koike and S.~Hirao at Matsushita Electric
Works, Ltd. for their support.
We thank all of the staff in Central Research Laboratory at Matsushita
Electric Works, Ltd. We thank the KEK machine shop members
for their help in constructing the test counter.
H.~Kawai is also acknowledged for stimulating discussions.

{}
\end{document}